\documentclass[[aps,pra,reprint,superscriptaddress,showpacs,letterpaper,longbibliography]{revtex4-1}

\usepackage[english]{babel}
\usepackage{ucs}
\usepackage[utf8x]{inputenc}
\usepackage{amsmath}
\usepackage{amsfonts}
\usepackage{amssymb}
\usepackage{graphicx}
\usepackage{bm}
\usepackage{bbm}
\usepackage{empheq}
\usepackage{color}

\definecolor{mjg}{rgb}{.08,.05,.8}

\definecolor{yyl}{rgb}{.8,.05,.08}

\newcommand{\delete}[1]{{}}

\begin{document}

\title{On-Chip Quantum Dot Light Source for Quantum State Readout}

\author{Y.-Y. Liu}
\author{J. Stehlik}
\author{X. Mi}
\author{T. Hartke}
\author{J. R. Petta}
\affiliation{Department of Physics, Princeton University, Princeton, New Jersey 08544, USA}

\date{\today}

\begin{abstract}
We use microwave radiation generated by a semiconductor double quantum dot (DQD) micromaser for charge state detection. A cavity is populated with $n_c \sim 6000$ photons by driving a current through an emitter DQD. These photons are used to sense the charge state of a target DQD that is located at the opposite end of the cavity. Charge dynamics in the target DQD influence the output power and emission frequency of the maser. Three different readout mechanisms are compared. The detection scheme requires no cavity input field and may potentially be used to improve the scalability of semiconductor and superconducting qubit readout technologies.
\end{abstract}


\maketitle

\section{Introduction}
Quantum state readout is a crucial component of any quantum computing architecture. For semiconductor quantum dots, charge state readout has been performed using quantum point contacts \cite{Field1993} and quantum dots \cite{Barthel2010} as detectors. Electronic spin states can also be resolved using spin-to-charge conversion, which relies on spin selective tunneling and sensitive charge state detection  \cite{Elzerman2004, Johnson2005}. To increase measurement bandwidths, conventional dc transport measurement approaches have to a large extent been replaced by rf and microwave frequency reflectometry setups \cite{Schoelkopf1998, Cottet2011, Petersson2010}. In particular, the circuit quantum electrodynamics (cQED) architecture allows for dispersive readout of superconducting qubits \cite{Blais2004, Wallraff2004, Sillanpaa2007, Reed2012}, as well as semiconductor charge and spin qubits \cite{Petersson2012, Delbecq2011, Frey2012, Mi2017, Bruhat2016, Stockklauser2017}.

Both rf-reflectometry and cQED measurement implementations rely on costly room temperature microwave sources, rf components, and coaxial lines that occupy a significant amount of space in a dilution refrigerator. As one scales to many qubits, the resource requirements will increase dramatically. Moreover, to suppress the room temperature microwave background, a typical attenuation of 60--70 dB is required in the coax lines connecting the signal generator to the quantum device ($\sim$ 10 mK). To reduce the experimental complexity, the source would ideally be isolated from the 300 K environment. 

Over the past 10 years it has been shown that a variety of voltage-biased quantum devices generate microwave frequency photons. For example, voltage-biased Cooper pair boxes and superconducting quantum intereference devices embedded in superconducting cavities have been shown to mase \cite{Astafiev2007, Cassidy2017}. Cavity-coupled semiconductor double quantum dots (DQD) can serve as an electrically tunable maser gain medium \cite{Stockklauser2015, Stehlik2016, Liu2017}. These devices are fabrication compatible with other qubits and they can be integrated on the same chip. 
It is therefore of interest to determine if these devices, which already operate at millikelvin temperatures, can be utilized as microwave frequency sources in quantum computing experiments \cite{Berg2014}.

\begin{figure}[t]
  \begin{center}
		\includegraphics[width=\columnwidth]{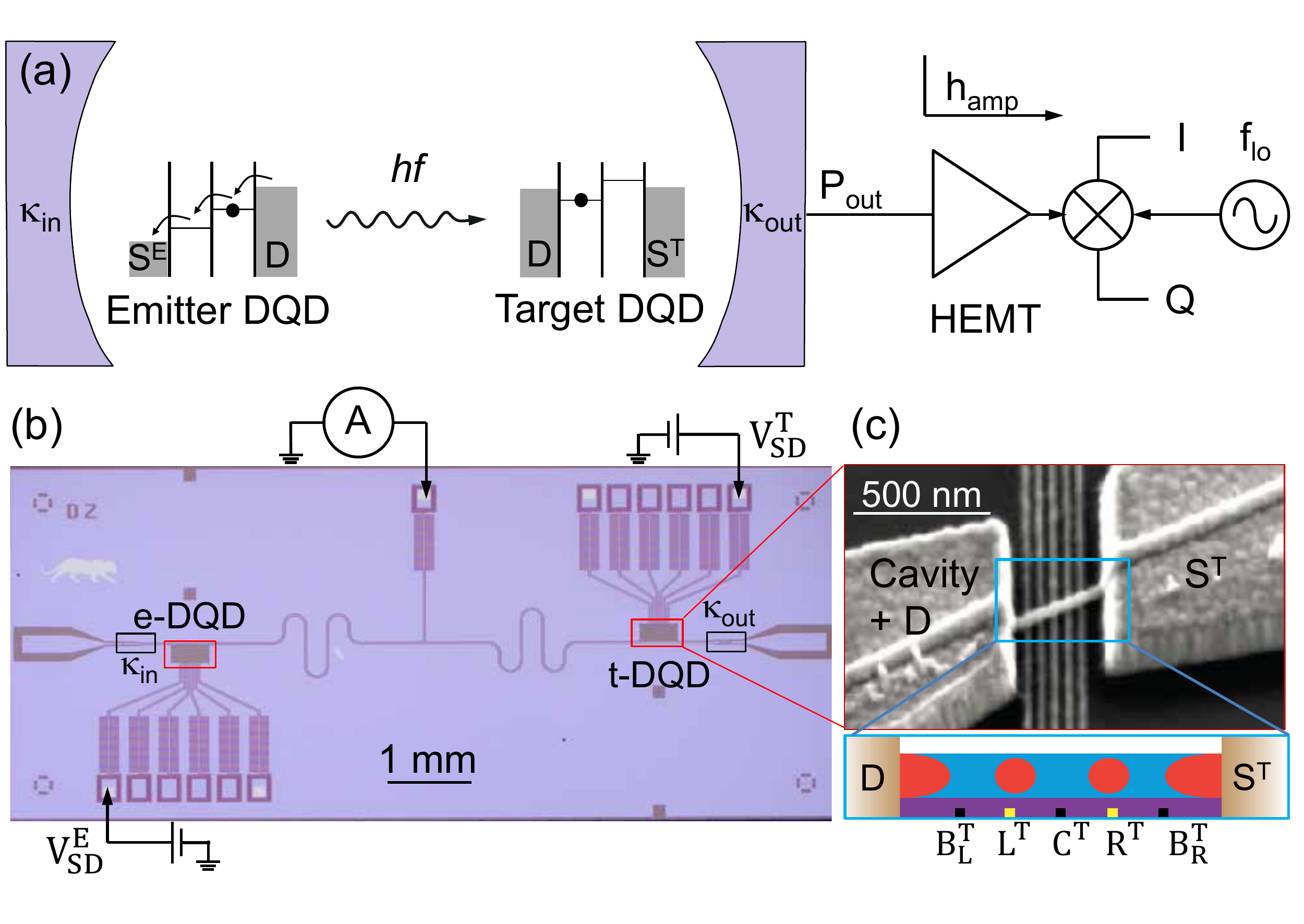}
  \caption{\label{Fig: scheme} (a) The experiment consists of a cavity containing two DQDs. A dc biased e-DQD mases and populates the cavity with microwave frequency photons. These photons are used to read out the charge state of the t-DQD. (b) Optical microscope image showing the cavity and the positions of the e-DQD and t-DQD. (c) Upper panel: Scanning electron micrograph of a nanowire DQD. Lower panel: The DQD confinement potential is defined using 5 bottom gates.}
  \end{center}
\end{figure}

In this Letter we show that microwave frequency photons generated by a cavity-coupled DQD can be used to sensitively readout the charge state of a second DQD that is located in the same cavity. A source-drain bias is applied across an emitter DQD (e-DQD) and results in above-threshold masing. The photons generated by the e-DQD are used to sense a target DQD (t-DQD) that is located in the same cavity. Charge dynamics in the t-DQD influence the maser emission, changing its output power and emission frequency, allowing for charge state readout.

We explore three different readout approaches. In the first approach the total power emitted by the cavity is measured and used to reconstruct the t-DQD charge stability diagram. In the second approach, the e-DQD emission frequency, which is dependent on the charge state of the t-DQD, is used to measure the t-DQD charge stability diagram by mapping out the maser emission frequency as a function of the t-DQD gate voltages. In the third approach, we measure the power emitted into a narrow band centered around the free-running maser frequency. Shifts in the emission frequency significantly change the narrow-band power, allowing us to detect t-DQD interdot charge transitions. While the target qubit in these experiments is a semiconductor DQD, it could in principle be replaced with any other cavity-coupled qubit, such as a superconducting transmon qubit. No external cavity drive is required in our experiments, further supporting the use of a DQD maser as a light source for quantum state readout in quantum computation architectures.

\section{Device Description}

Figure\ 1(a) captures the important elements of our experimental setup. The e-DQD and t-DQD are coupled to a microwave cavity. To serve as a microwave source, the e-DQD is source-drain biased, which results in single electron tunneling events and microwave frequency photon emission into the cavity mode \cite{Liu2017}. These cavity photons are used to sense the charge state of the t-DQD.

An optical micrograph of the device is shown in Fig.~\ref{Fig: scheme}(b). The cavity consists of a $\lambda/2$ superconducting resonator with resonance frequency $f_{\rm c} = 7596$ MHz and linewidth $\kappa_{\rm tot}/2\pi$ = 1.77 MHz. The cavity is designed to have output (input) coupling rates $\kappa_{\rm out}/2\pi = $ 0.8 MHz ($\kappa_{\rm in}/2\pi = $ 0.04 MHz). The InAs nanowire DQDs are located at opposite ends of the cavity near the voltage anti-nodes. The confinement potential of the t-DQD is created by voltage biasing the five bottom gates labeled as $\rm{B^T_L}$, $\rm{L^T}$, ${\rm C^T}$, ${\rm R^T}$, and $\rm{B^T_R}$ in Fig.\ 1(c). Independent control of the e-DQD is achieved using a separate set of bottom gates. We further define source and drain ohmic contacts using electron beam lithography. In contrast with our previous work, the source contacts to the e-DQD and t-DQD are electrically decoupled such that the source-drain bias voltages can be independently controlled \cite{Liu2015}. The drain electrode of each DQD is connected to the microwave resonator. Coupling between a charge trapped in the DQD confinement potential and the electric field of the cavity leads to an electric dipole interaction with strength $g_c/2\pi \approx 30$ -- $40$ MHz \cite{Stehlik2016}.

\section{Characterization of the Emitter Double Quantum Dot}

We first characterize the microwave field emitted by the e-DQD. For these measurements, the t-DQD source drain bias $V^{\rm T}_{\rm SD} = 0$ and the gates are tuned such that the t-DQD is in Coulomb blockade, where the charge number is fixed. In this configuration the t-DQD will not have any impact on the cavity field. The e-DQD is configured to emit photons by applying a finite source drain bias $V^{\rm E}_{\rm SD} = 2$ mV, which results in single electron tunneling events. The interdot charge transition leads to photon emission and, in a high quality factor cavity, a transition to a masing state \cite{Liu2017}. 

\begin{figure}[t]
  \begin{center}
		\includegraphics[width=\columnwidth]{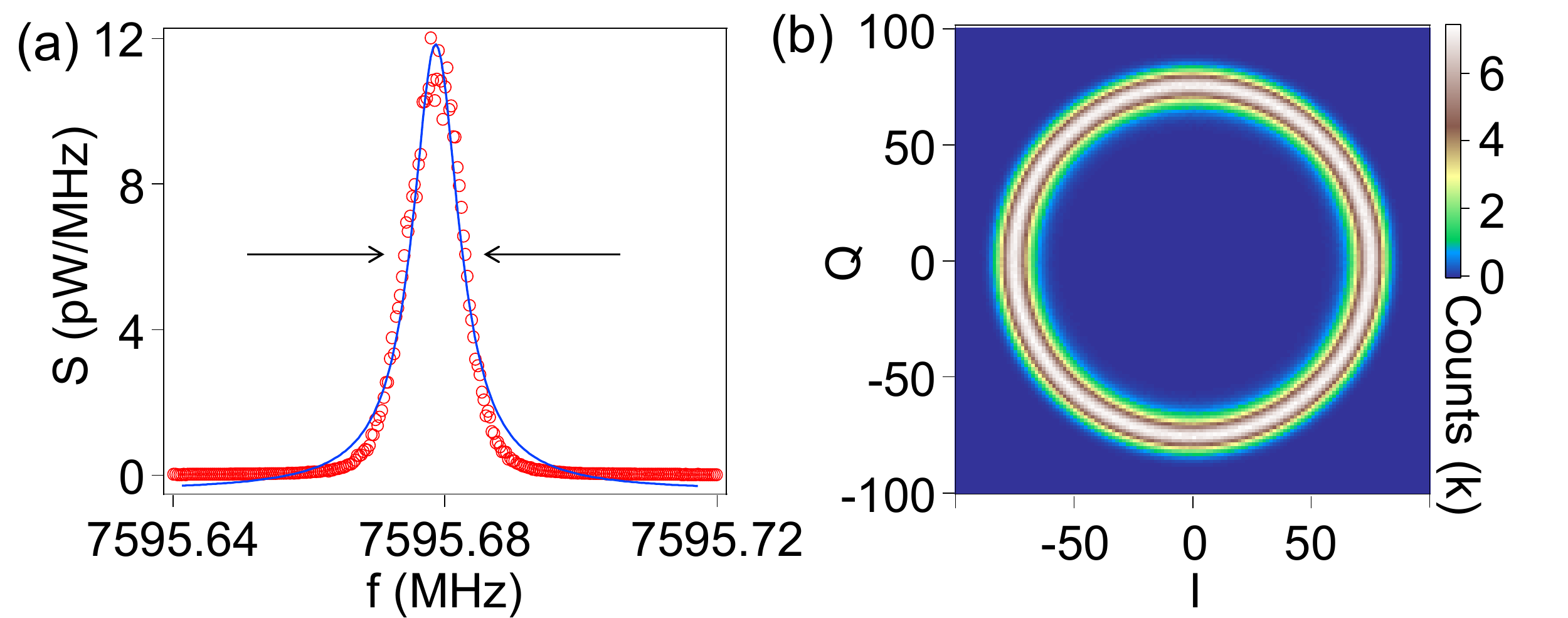}
  \caption{\label{Fig: LDQD emission} (a) Power spectral density of the radiation emitted by the e-DQD (circles) and a fit to a Lorentzian (solid line) with FWHM = 8 kHz. (b) The $IQ$ histogram of the output field is consistent with a coherent source.}
  \end{center}
\end{figure}

To measure the emitted radiation, the cavity output field is amplified using a high electron mobility transistor (HEMT) amplifier and detected with a spectrum analyzer. Figure\ \ref{Fig: LDQD emission}(a) shows the power spectral density $S(f)$ of the radiation emitted from the cavity, along with a fit to a Lorentzian. The best fit parameters yield the emission frequency  $f_{\rm e} = 7595.68$ MHz and FWHM = 8~kHz. We obtain a total output power $P_{\rm out} =$ 0.16 pW by integrating over $S(f)$. The emission power yields an intra-cavity photon number $n_{\rm c} = P_{\rm out}/(hf_e\kappa_{\rm out})\approx 6000$ given $\kappa_{\rm out}/2\pi = 0.8$ MHz. The FWHM is 200 times narrower than the bare cavity linewidth, which is suggestive of masing.

The output field can be examined in more detail by measuring ($I$,$Q$) histograms. To acquire the histograms, the cavity output field is first amplified with a HEMT and then demodulated into the in-phase ($I$) and quadrature-phase ($Q$) components by a local reference set to a frequency $f_{\rm lo} = f_{\rm c}$ \cite{Liu2015}. Figure~\ref{Fig: LDQD emission}(b) shows an ($I$,$Q$) histogram obtained by accumulating $1.7\times10^7$  $(I,Q)$ samples at a rate of 12.3 MHz. The histogram has a ring shape that is consistent with coherent emission \cite{Liu2017}. Combined, these data sets show that the voltage-biased e-DQD can serve as a coherent source that populates the cavity with approximately 6000 photons. These photons may be used to read out the charge state of the t-DQD, as will be demonstrated in the following sections of the paper.

\section{Target Double Quantum Dot Charge State Detection}

In this section we compare several different approaches for measuring the charge stability diagram of the t-DQD. We first measure the stability diagram using standard cavity input-output readout, where an external tone is used to populate the cavity with photons. These data are then compared with charge stability diagrams that are obtained by measuring the total power emitted from the cavity when it is populated with e-DQD photons. Two additional transduction methods are examined that are based on the effect that charge dynamics in the t-DQD have on the emission properties of the e-DQD. Specifically, we show that the t-DQD charge stability diagram can be reconstructed by measuring the emission frequency of the e-DQD and the narrow band power emitted by the e-DQD.

\subsection{Charge state readout through measurements of the cavity transmission}

\begin{figure}[t]
	\begin{center}
		\includegraphics[width=\columnwidth]{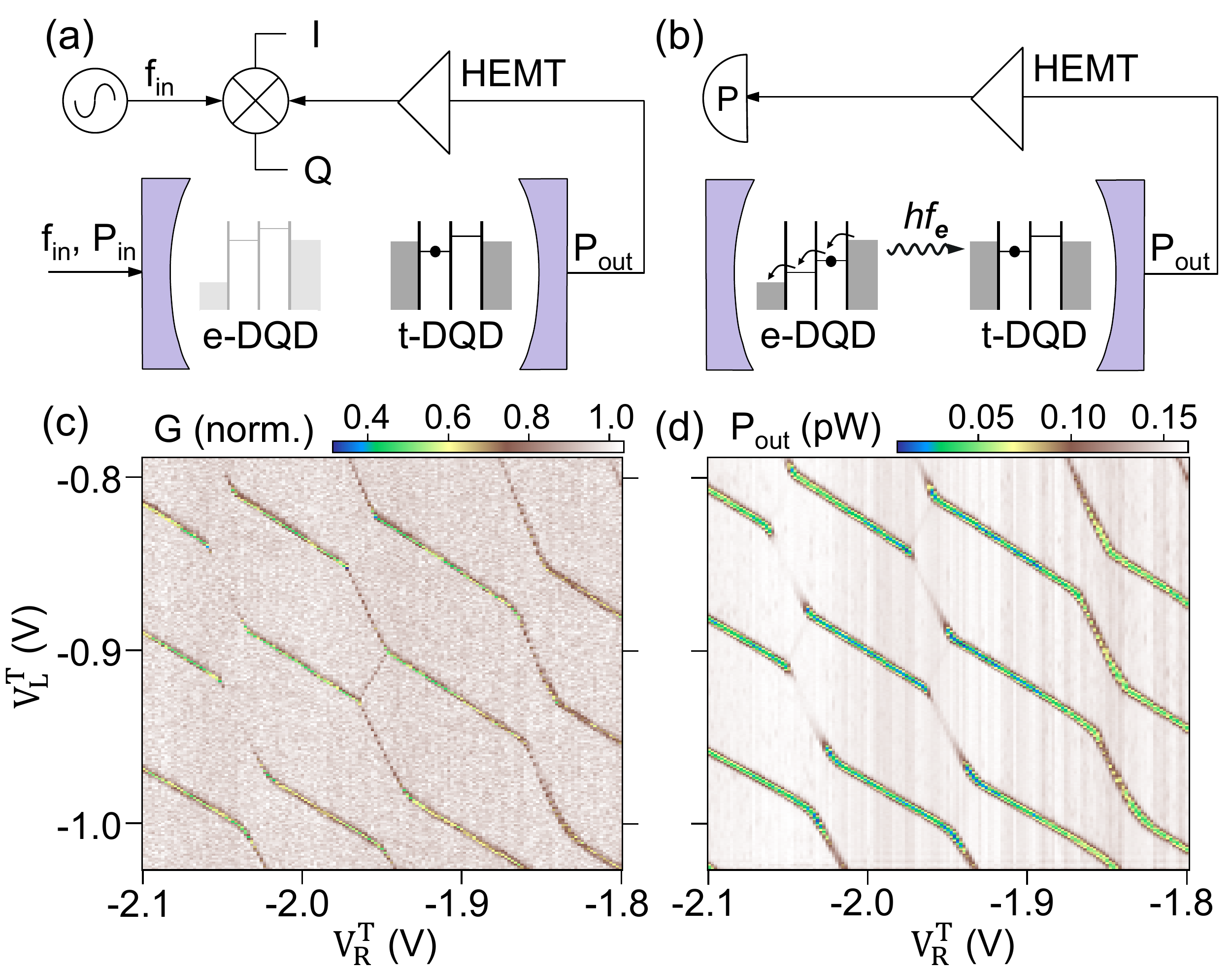}
		\caption{\label{Fig: total power sensing} (a) Circuit used to measure the t-DQD charge stability diagram by driving the cavity with a weak input tone while the e-DQD is in Coulomb blockade. (b) Photons emitted from the e-DQD can be used to measure the charge stability diagram of the t-DQD in the absence of a cavity input tone. (c) Gain, $G$, measured with method shown in (a) as a function of $V^{\rm T}_{\rm L}$ and $V^{\rm T}_{\rm R}$, revealing the t-DQD charge stability diagram. (d) $P_{\rm out}$ measured with method shown in (b) as a function of  $V^{\rm T}_{\rm L}$ and $V^{\rm T}_{\rm R}$ also reveals the t-DQD charge stability diagram. }
	\end{center}	
\end{figure}

The conventional cavity input-output readout approach is illustrated in Fig.\ \ref{Fig: total power sensing}(a). Here the cavity is driven by an input tone of frequency $f_{\rm in}$ and power $P_{\rm in} \approx -112$~dBm that results in approximately $n_{\rm c}\approx10$ intra-cavity photons. The resulting cavity output is amplified with a HEMT and demodulated by a local reference having a frequency $f_{\rm lo} = f_{\rm in}$. Both the phase shift $\Delta \phi$ and power gain $G = CP_{\rm out}/P_{\rm in}$ can be extracted from the cavity transmission. Here the constant $C$ is set such that $G = 1$ with $f_{\rm in} = f_{\rm c}$ and both DQDs in Coulomb blockade \cite{Liu2015, Stehlik2016}. Figure\ \ref{Fig: total power sensing}(c) plots $G$ as a function of the t-DQD gate voltages with $f_{\rm in} = f_{\rm c}$ and $V^{\rm T}_{\rm SD} = 0$. For this data set the e-DQD is in idle mode, with $V^{\rm E}_{\rm SD} = 0$ and the gate voltages tuned to Coulomb blockade. These measurements reveal the t-DQD charge stability diagram, consistent with previous measurements of cavity-coupled InAs nanowire DQDs \cite{Stehlik2016}.

\subsection{Charge state readout through measurements of the total cavity output power}

To make a comparison with cavity input-output readout we now turn off the cavity input tone and configure the e-DQD in the ``on state," such that it is emitting coherent radiation as shown in Fig.\ 2. We then measure the output power $P_{\rm out}$ and plot it as a function of $V^{\rm T}_{\rm L}$ and $V^{\rm T}_{\rm R}$ in Fig.~\ref{Fig: total power sensing}(d). Writing the cavity output field complex amplitude as $\alpha = I + iQ$, $P_{\rm out}$ is determined from measurements of $\langle \alpha^*\alpha\rangle = \langle I^2+Q^2\rangle$. The $(I,Q)$ data are processed using a digital filter of 2.6 MHz bandwidth that covers the entire cavity linewidth and therefore $ \langle I^2+Q^2\rangle$ captures the total emitted power \cite{EichlerPRA2012}. The scenario is equivalent to a power meter measuring over a wide bandwidth as illustrated in Fig.\ \ref{Fig: total power sensing}(b). The data in Fig.~\ref{Fig: total power sensing}(d) show that measurements of $P_{\rm out}$ can be used to extract the t-DQD charge stability diagram.


\subsection{Impact of charge dynamics in the t-DQD on the emission properties of the e-DQD}

\begin{figure*}[t]
	\begin{center}
		\includegraphics[width=2\columnwidth]{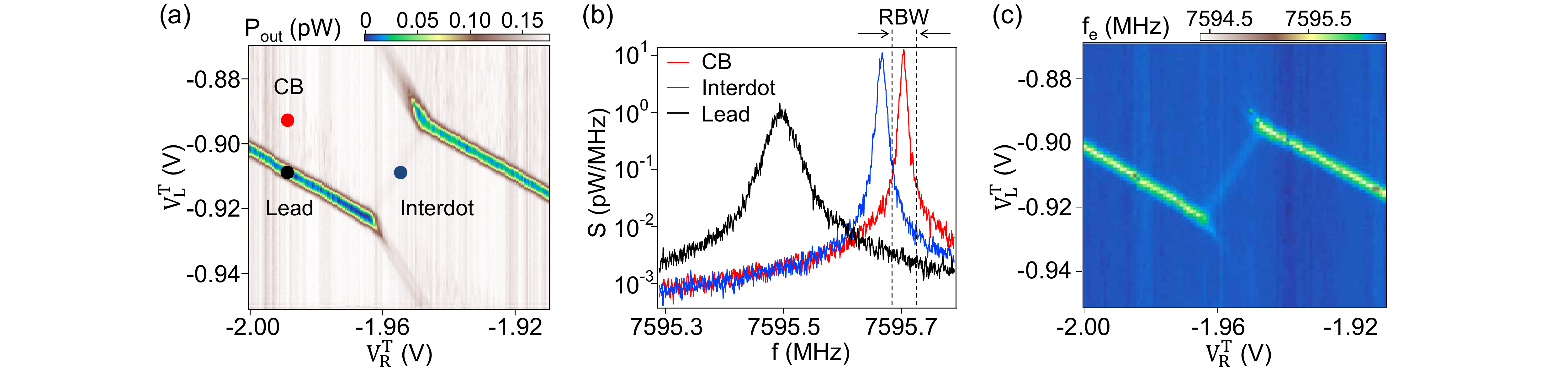}
		\caption{\label{Fig: spectrum dependance} (a) Total output power $P_{\rm out}$ as a function of $V^{\rm T}_{\rm L}$ and $V^{\rm T}_{\rm R}$ near a t-DQD interdot charge transition. Here the e-DQD emission is used to populate the cavity with measurement photons. (b) $S(f)$ measured with the t-DQD in Coulomb blockade (CB), at an interdot transition, and at a left dot charge transition (Lead). The dashed lines indicate the frequency window that the emission is integrated over in the narrowband measurement approach. (c) $f_{\rm e}$ as a function of $V^{\rm T}_{\rm L}$ and $V^{\rm T}_{\rm R}$. In these data both single dot and interdot charge transitions are visible.}
	\end{center}
\end{figure*}

We now more carefully examine the readout mechanism by studying the effect that the t-DQD charge configuration has on the emission properties of the e-DQD. Figure\ \ref{Fig: spectrum dependance}(a) shows a high resolution measurement of $P_{\rm out}$ near one of the t-DQD interdot charge transitions in Fig.\ \ref{Fig: total power sensing}(d). These data were acquired in the absence of a cavity input tone and with the e-DQD emitting photons. The left dot and right dot charge transitions are visible in these data, while the visibility of the interdot charge transition is significantly less than in the data shown in Fig.\ 3(c). 

To better understand what sets the visibility of the charge transitions in these data, we measure $S(f)$ of the emitted radiation with the gate voltages of the t-DQD tuned to different regions of the t-DQD charge stability diagram. Figure\ \ref{Fig: spectrum dependance}(b) shows measurements of $S(f)$ with the t-DQD configured in Coulomb blockade, at the interdot charge transition, and at a left dot charge transition. With the t-DQD configured in Coulomb blockade the emission peak in $S(f)$ is centered at $f_{\rm e}^0$ = 7595.68~MHz. When the t-DQD is configured to a left dot charge transition, the emission peak shifts down in frequency by 214 kHz, the peak power is reduced by approximately a factor of 10, and the peak in $S(f)$ is significantly broader. In comparison, with the t-DQD configured at the interdot charge transition the emission peak is only shifted down in frequency by 37 kHz. The emission peak has a height and width that is comparable to the data acquired with the t-DQD in Coulomb blockade. Therefore, it is difficult to resolve the interdot charge transitions in measurements of the total emitted power $P_{\rm out}$. However, since the emission peak shifts by an amount that is much greater than the FWHM $\sim 8$ kHz of the emission peak, measurements of the emission frequency may be used to reconstruct the t-DQD charge stability diagram.

By fitting $S(f)$ to a Lorentzian at every point in the t-DQD charge stability diagram we can extract $f_{\rm e}$ as a function of $V^{\rm T}_{\rm L}$ and $V^{\rm T}_{\rm R}$. A plot of the extracted $f_e$ is shown in Fig.\ \ref{Fig: spectrum dependance}(c) and is much more sensitive to the interdot charge transition. Therefore a measurement of $f_{\rm e}$ can in principle be used to readout the device. The approach is similar to cQED readout of transmon qubits, where the state-dependent dispersive shift of the cavity is used for readout \cite{Wallraff2004, Blais2004}.

It is important to note here that in general we do not know the phase of the maser emission, and that previous work showed that the coherence time of the maser is only on the order of 10 $\mu$s \cite{Liu2015}. Even with a long coherence time, $f_e = f_e(V^{\rm T}_{\rm L}, V^{\rm T}_{\rm R})$ is t-DQD dependent and thus the phase shift in the maser output $\Delta \phi(t) = \int \Delta f_e(t) dt$ where $f_e(t) =f_e(V^{\rm T}_{\rm L}(t), V^{\rm T}_{\rm R}(t))$ is dependent on the ``path" of $(V^{\rm T}_{\rm L}(t), V^{\rm T}_{\rm R}(t))$. $\Delta \phi$ is then not a well defined variable. Therefore we cannot simply measure the dispersive shift $\Delta \phi$, as is commonly achieved with phase-sensitive measurement approaches in cQED.

\subsection{Charge state readout through narrow-band measurements of the total cavity output power}

\begin{figure}[t]
	\begin{center}
		\includegraphics[width=\columnwidth]{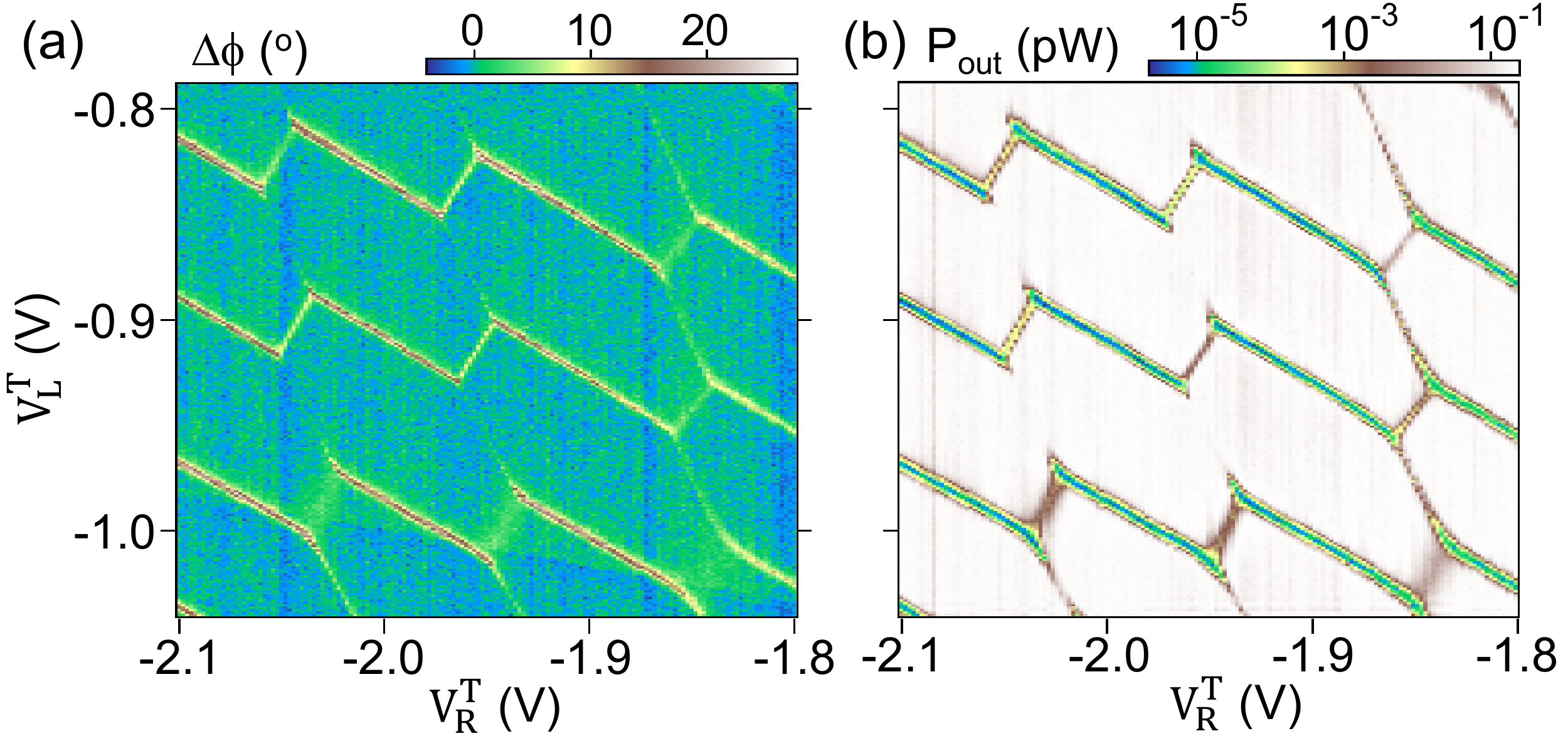}
		\caption{\label{Fig: narrow rbw readout} (a) Cavity phase shift $\Delta \phi$ measured in response to a weak cavity input tone as a function of $V^{\rm T}_{\rm L}$ and $V^{\rm T}_{\rm R}$. (b) $P_{\rm out}$ integrated within a 30 kHz RBW as a function of $V^{\rm T}_{\rm L}$ and $V^{\rm T}_{\rm R}$. The integrated power in this narrow bandwidth is sensitive to changes in $f_{\rm e}$.}
	\end{center}	
\end{figure}

The previous section demonstrated that measurements of the emission frequency $f_{\rm e}$ can be used to reconstruct the t-DQD charge stability diagram. However, extracting $f_{\rm e}$ from measurements of $S(f)$ is too time consuming (3--4 seconds per spectrum) to allow for efficient charge state readout. The challenge of devising a practical measurement that quickly extracts the state-dependent frequency shift has been solved in the standard readout schemes. For example, in cQED systems, state dependent shifts in the resonance frequency of the cavity can be measured by driving the cavity with a weak input tone at $f_{\rm in} = f_{\rm c}$ and detecting the large phase shift $\Delta \phi = \arctan(I,Q)$ of the cavity output field using heterodyne demodulation techniques. As a demonstration of the standard readout approach, Fig.~\ref{Fig: narrow rbw readout}(a) plots the phase shift $\Delta \phi$ as a function of $V^{\rm T}_{\rm L}$ and $V^{\rm T}_{\rm R}$. Single dot transitions associated with the left and right dots, as well as the interdot charge transition, are clearly visible in the phase response.

Phase readout is not feasible when e-DQD emission is used to populate the cavity with photons since $f_{\rm e}$ exhibits fluctuations that randomize the phase. Moreover, since $f_{\rm e}$ is a quantity that depends on the t-DQD configuration the phase shift is not a well-defined quantity.  Instead, a quantity analogous to the phase shift can be measured for fast readout. The emission spectrum [Fig.\ 4(b)] shifts in response to the charge state of the t-DQD, allowing us to simply measure the output power $P_{\rm out}$ within a narrow resolution bandwidth (RBW), as schematically illustrated in Fig.\ \ref{Fig: spectrum dependance}(b). The frequency range over which the power is integrated $f_{\rm e}^0 \pm$ RBW$/2$ should be smaller than the expected state-dependent shift in $f_{\rm e}$, yet large enough to tolerate the drift in $f_{\rm e}$ caused by charge fluctuations in the emitter \cite{LiuPRA2015, Liu2017}. We operate with RBW $>$ FWHM of the e-DQD emission spectrum to tolerate the drift in $f_{\rm e}$, and RBW $<\left|f_{\rm e}-f_{\rm e}^0\right|$ to allow sensitivity to changes in the emission spectrum due to the t-DQD charge state. Figure \ref{Fig: narrow rbw readout} (b) shows the output power $P_{\rm out}$ measured around $f_{\rm e}^0$ = 7595.68 MHz with a 30 kHz RBW. The state-dependent shift in the emission center frequency at a t-DQD interdot charge transition leads to a factor of 100 change in $P_{\rm out}$ within the measured bandwidth. 

\section{Summary and Outlook}

In summary, we have shown that a voltage-biased DQD can be used as a light source for qubit readout in the cQED architecture. Readout based on measurements of the total output power, emission center frequency, and narrow-band output power were compared. While the total output power is sensitive to single dot charge transitions, it does not have sufficient sensitivity to resolve interdot charge transitions. Measurements of the emission center frequency reveal both single dot and interdot charge transitions, but this approach is slow and not well-suited for single shot readout. The narrow-band power measurement approach yields high sensitivity to both single dot and interdot charge transitions. 

In some applications, it may be desirable to place the e-DQD in a separate cavity. In the masing state, the e-DQD generates a large intra-cavity photon number $n_{\rm c} \sim 6000$, which may cause saturation effects and broaden the linewidth of the target transition. Separating the emitter from the target qubit would more easily allow the emitted field to be attenuated. Lastly, previous work has shown that the maser can be switched on and off rapidly \cite{Liu2017}. A switchable maser could be turned off during quantum control sequences and then rapidly activated for high power readout of the qubit state \cite{Reed2010}. We hope that this study will motivate further applications of nanoscale emitters in quantum computing readout architectures.

\begin{acknowledgments}
We thank M. J. Gullans and J. M. Taylor for helpful discussions and acknowledge support from the Packard Foundation, the National Science Foundation Grant No.\ DMR-1409556, and the Gordon and Betty Moore Foundation’s EPiQS Initiative through Grant GBMF4535. Devices were fabricated in the Princeton University Quantum Device Nanofabrication Laboratory.
\end{acknowledgments}
\bibliographystyle{apsrev_lyy2017}
\bibliography{lightbulb_v2}

\begin{thebibliography}{27}
\expandafter\ifx\csname natexlab\endcsname\relax\def\natexlab#1{#1}\fi
\expandafter\ifx\csname bibnamefont\endcsname\relax
  \def\bibnamefont#1{#1}\fi
\expandafter\ifx\csname bibfnamefont\endcsname\relax
  \def\bibfnamefont#1{#1}\fi
\expandafter\ifx\csname citenamefont\endcsname\relax
  \def\citenamefont#1{#1}\fi
\expandafter\ifx\csname url\endcsname\relax
  \def\url#1{\texttt{#1}}\fi
\expandafter\ifx\csname urlprefix\endcsname\relax\def\urlprefix{URL }\fi
\providecommand{\bibinfo}[2]{#2}
\providecommand{\eprint}[2][]{\url{#2}}

\bibitem[{\citenamefont{Field et~al.}(1993)\citenamefont{Field, Smith, Pepper,
  Ritchie, Frost, Jones, and Hasko}}]{Field1993}
\bibinfo{author}{\bibfnamefont{M.}~\bibnamefont{Field}},
  \bibinfo{author}{\bibfnamefont{C.~G.} \bibnamefont{Smith}},
  \bibinfo{author}{\bibfnamefont{M.}~\bibnamefont{Pepper}},
  \bibinfo{author}{\bibfnamefont{D.~A.} \bibnamefont{Ritchie}},
  \bibinfo{author}{\bibfnamefont{J.~E.~F.} \bibnamefont{Frost}},
  \bibinfo{author}{\bibfnamefont{G.~A.~C.} \bibnamefont{Jones}},
  \bibnamefont{and} \bibinfo{author}{\bibfnamefont{D.~G.} \bibnamefont{Hasko}},
  Measurements of Coulomb blockade with a noninvasive voltage probe,
  \bibinfo{journal}{Phys. Rev. Lett.} \textbf{\bibinfo{volume}{70}},
  \bibinfo{pages}{1311} (\bibinfo{year}{1993}).

\bibitem[{\citenamefont{Barthel et~al.}(2010)\citenamefont{Barthel,
  Kj\ae{}rgaard, Medford, Stopa, Marcus, Hanson, and Gossard}}]{Barthel2010}
\bibinfo{author}{\bibfnamefont{C.}~\bibnamefont{Barthel}},
  \bibinfo{author}{\bibfnamefont{M.}~\bibnamefont{Kj\ae{}rgaard}},
  \bibinfo{author}{\bibfnamefont{J.}~\bibnamefont{Medford}},
  \bibinfo{author}{\bibfnamefont{M.}~\bibnamefont{Stopa}},
  \bibinfo{author}{\bibfnamefont{C.~M.} \bibnamefont{Marcus}},
  \bibinfo{author}{\bibfnamefont{M.~P.} \bibnamefont{Hanson}},
  \bibnamefont{and} \bibinfo{author}{\bibfnamefont{A.~C.}
  \bibnamefont{Gossard}}, Fast sensing of double-dot charge arrangement and
  spin state with a radio-frequency sensor quantum dot, \bibinfo{journal}{Phys.
  Rev. B} \textbf{\bibinfo{volume}{81}}, \bibinfo{pages}{161308}
  (\bibinfo{year}{2010}).

\bibitem[{\citenamefont{Elzerman et~al.}(2004)\citenamefont{Elzerman, Hanson,
  Willems~van Beveren, Witkamp, Vandersypen, and Kouwenhoven}}]{Elzerman2004}
\bibinfo{author}{\bibfnamefont{J.~M.} \bibnamefont{Elzerman}},
  \bibinfo{author}{\bibfnamefont{R.}~\bibnamefont{Hanson}},
  \bibinfo{author}{\bibfnamefont{L.~H.} \bibnamefont{Willems~van Beveren}},
  \bibinfo{author}{\bibfnamefont{B.}~\bibnamefont{Witkamp}},
  \bibinfo{author}{\bibfnamefont{L.~M.~K.} \bibnamefont{Vandersypen}},
  \bibnamefont{and} \bibinfo{author}{\bibfnamefont{L.~P.}
  \bibnamefont{Kouwenhoven}}, Single-shot read-out of an individual electron
  spin in a quantum dot, \bibinfo{journal}{Nature (London)}
  \textbf{\bibinfo{volume}{430}}, \bibinfo{pages}{431} (\bibinfo{year}{2004}).

\bibitem[{\citenamefont{Johnson et~al.}(2005)\citenamefont{Johnson, Petta,
  Marcus, Hanson, and Gossard}}]{Johnson2005}
\bibinfo{author}{\bibfnamefont{A.~C.} \bibnamefont{Johnson}},
  \bibinfo{author}{\bibfnamefont{J.~R.} \bibnamefont{Petta}},
  \bibinfo{author}{\bibfnamefont{C.~M.} \bibnamefont{Marcus}},
  \bibinfo{author}{\bibfnamefont{M.~P.} \bibnamefont{Hanson}},
  \bibnamefont{and} \bibinfo{author}{\bibfnamefont{A.~C.}
  \bibnamefont{Gossard}}, Singlet-triplet spin blockade and charge sensing in a
  few-electron double quantum dot, \bibinfo{journal}{Phys. Rev. B}
  \textbf{\bibinfo{volume}{72}}, \bibinfo{pages}{165308}
  (\bibinfo{year}{2005}).

\bibitem[{\citenamefont{Schoelkopf et~al.}(1998)\citenamefont{Schoelkopf,
  Wahlgren, Kozhevnikov, Delsing, and Prober}}]{Schoelkopf1998}
\bibinfo{author}{\bibfnamefont{R.~J.} \bibnamefont{Schoelkopf}},
  \bibinfo{author}{\bibfnamefont{P.}~\bibnamefont{Wahlgren}},
  \bibinfo{author}{\bibfnamefont{A.~A.} \bibnamefont{Kozhevnikov}},
  \bibinfo{author}{\bibfnamefont{P.}~\bibnamefont{Delsing}}, \bibnamefont{and}
  \bibinfo{author}{\bibfnamefont{D.~E.} \bibnamefont{Prober}}, The
  Radio-Frequency Single-Electron Transistor (RF-SET): A Fast and
  Ultrasensitive Electrometer, \bibinfo{journal}{Science}
  \textbf{\bibinfo{volume}{280}}, \bibinfo{pages}{1238} (\bibinfo{year}{1998}).

\bibitem[{\citenamefont{Cottet et~al.}(2011)\citenamefont{Cottet, Mora, and
  Kontos}}]{Cottet2011}
\bibinfo{author}{\bibfnamefont{A.}~\bibnamefont{Cottet}},
  \bibinfo{author}{\bibfnamefont{C.}~\bibnamefont{Mora}}, \bibnamefont{and}
  \bibinfo{author}{\bibfnamefont{T.}~\bibnamefont{Kontos}}, Mesoscopic
  admittance of a double quantum dot, \bibinfo{journal}{Phys. Rev. B}
  \textbf{\bibinfo{volume}{83}}, \bibinfo{pages}{121311}
  (\bibinfo{year}{2011}).

\bibitem[{\citenamefont{Petersson et~al.}(2010)\citenamefont{Petersson, Smith,
  Anderson, Atkinson, Jones, and Ritchie}}]{Petersson2010}
\bibinfo{author}{\bibfnamefont{K.~D.} \bibnamefont{Petersson}},
  \bibinfo{author}{\bibfnamefont{C.~G.} \bibnamefont{Smith}},
  \bibinfo{author}{\bibfnamefont{D.}~\bibnamefont{Anderson}},
  \bibinfo{author}{\bibfnamefont{P.}~\bibnamefont{Atkinson}},
  \bibinfo{author}{\bibfnamefont{G.~A.~C.} \bibnamefont{Jones}},
  \bibnamefont{and} \bibinfo{author}{\bibfnamefont{D.~A.}
  \bibnamefont{Ritchie}}, Charge and Spin State Readout of a Double Quantum Dot
  Coupled to a Resonator, \bibinfo{journal}{Nano Lett.}
  \textbf{\bibinfo{volume}{10}}, \bibinfo{pages}{2789} (\bibinfo{year}{2010}).

\bibitem[{\citenamefont{Blais et~al.}(2004)\citenamefont{Blais, Huang,
  Wallraff, Girvin, and Schoelkopf}}]{Blais2004}
\bibinfo{author}{\bibfnamefont{A.}~\bibnamefont{Blais}},
  \bibinfo{author}{\bibfnamefont{R.-S.} \bibnamefont{Huang}},
  \bibinfo{author}{\bibfnamefont{A.}~\bibnamefont{Wallraff}},
  \bibinfo{author}{\bibfnamefont{S.~M.} \bibnamefont{Girvin}},
  \bibnamefont{and} \bibinfo{author}{\bibfnamefont{R.~J.}
  \bibnamefont{Schoelkopf}}, Cavity Quantum Electrodynamics for Superconducting
  Electrical Circuits: An Architecture for Quantum Computation,
  \bibinfo{journal}{Phys. Rev. A} \textbf{\bibinfo{volume}{69}},
  \bibinfo{pages}{062320} (\bibinfo{year}{2004}).

\bibitem[{\citenamefont{Wallraff et~al.}(2004)\citenamefont{Wallraff, Schuster,
  Blais, Frunzio, Huang, Majer, Kumar, Girvin, and Schoelkopf}}]{Wallraff2004}
\bibinfo{author}{\bibfnamefont{A.}~\bibnamefont{Wallraff}},
  \bibinfo{author}{\bibfnamefont{D.~I.} \bibnamefont{Schuster}},
  \bibinfo{author}{\bibfnamefont{A.}~\bibnamefont{Blais}},
  \bibinfo{author}{\bibfnamefont{L.}~\bibnamefont{Frunzio}},
  \bibinfo{author}{\bibfnamefont{R.-S.} \bibnamefont{Huang}},
  \bibinfo{author}{\bibfnamefont{J.}~\bibnamefont{Majer}},
  \bibinfo{author}{\bibfnamefont{S.}~\bibnamefont{Kumar}},
  \bibinfo{author}{\bibfnamefont{S.~M.} \bibnamefont{Girvin}},
  \bibnamefont{and} \bibinfo{author}{\bibfnamefont{R.~J.}
  \bibnamefont{Schoelkopf}}, Strong Coupling of a Single Photon to a
  Superconducting Qubit Using Circuit Quantum Electrodynamics,
  \bibinfo{journal}{Nature (London)} \textbf{\bibinfo{volume}{431}},
  \bibinfo{pages}{162} (\bibinfo{year}{2004}).

\bibitem[{\citenamefont{Sillanpaa et~al.}(2007)\citenamefont{Sillanpaa, Park,
  and Simmonds}}]{Sillanpaa2007}
\bibinfo{author}{\bibfnamefont{M.~A.} \bibnamefont{Sillanpaa}},
  \bibinfo{author}{\bibfnamefont{J.~I.} \bibnamefont{Park}}, \bibnamefont{and}
  \bibinfo{author}{\bibfnamefont{R.~W.} \bibnamefont{Simmonds}}, Coherent
  quantum state storage and transfer between two phase qubits via a resonant
  cavity, \bibinfo{journal}{Nature (London)} \textbf{\bibinfo{volume}{449}},
  \bibinfo{pages}{438} (\bibinfo{year}{2007}).

\bibitem[{\citenamefont{Reed et~al.}(2012)\citenamefont{Reed, DiCarlo, Nigg,
  Sun, Frunzio, Girvin, and Schoelkopf}}]{Reed2012}
\bibinfo{author}{\bibfnamefont{M.~D.} \bibnamefont{Reed}},
  \bibinfo{author}{\bibfnamefont{L.}~\bibnamefont{DiCarlo}},
  \bibinfo{author}{\bibfnamefont{S.~E.} \bibnamefont{Nigg}},
  \bibinfo{author}{\bibfnamefont{L.}~\bibnamefont{Sun}},
  \bibinfo{author}{\bibfnamefont{L.}~\bibnamefont{Frunzio}},
  \bibinfo{author}{\bibfnamefont{S.~M.} \bibnamefont{Girvin}},
  \bibnamefont{and} \bibinfo{author}{\bibfnamefont{R.~J.}
  \bibnamefont{Schoelkopf}}, Realization of three-qubit quantum error
  correction with superconducting circuits, \bibinfo{journal}{Nature (London)}
  \textbf{\bibinfo{volume}{482}}, \bibinfo{pages}{382} (\bibinfo{year}{2012}).

\bibitem[{\citenamefont{Petersson et~al.}(2012)\citenamefont{Petersson, McFaul,
  Schroer, Jung, Taylor, Houck, and Petta}}]{Petersson2012}
\bibinfo{author}{\bibfnamefont{K.~D.} \bibnamefont{Petersson}},
  \bibinfo{author}{\bibfnamefont{L.~W.} \bibnamefont{McFaul}},
  \bibinfo{author}{\bibfnamefont{M.~D.} \bibnamefont{Schroer}},
  \bibinfo{author}{\bibfnamefont{M.}~\bibnamefont{Jung}},
  \bibinfo{author}{\bibfnamefont{J.~M.} \bibnamefont{Taylor}},
  \bibinfo{author}{\bibfnamefont{A.~A.} \bibnamefont{Houck}}, \bibnamefont{and}
  \bibinfo{author}{\bibfnamefont{J.~R.} \bibnamefont{Petta}}, Circuit Quantum
  Electrodynamics with a Spin Qubit, \bibinfo{journal}{Nature (London)}
  \textbf{\bibinfo{volume}{490}}, \bibinfo{pages}{380} (\bibinfo{year}{2012}).

\bibitem[{\citenamefont{Delbecq et~al.}(2011)\citenamefont{Delbecq, Schmitt,
  Parmentier, Roch, Viennot, F\`eve, Huard, Mora, Cottet, and
  Kontos}}]{Delbecq2011}
\bibinfo{author}{\bibfnamefont{M.~R.} \bibnamefont{Delbecq}},
  \bibinfo{author}{\bibfnamefont{V.}~\bibnamefont{Schmitt}},
  \bibinfo{author}{\bibfnamefont{F.~D.} \bibnamefont{Parmentier}},
  \bibinfo{author}{\bibfnamefont{N.}~\bibnamefont{Roch}},
  \bibinfo{author}{\bibfnamefont{J.~J.} \bibnamefont{Viennot}},
  \bibinfo{author}{\bibfnamefont{G.}~\bibnamefont{F\`eve}},
  \bibinfo{author}{\bibfnamefont{B.}~\bibnamefont{Huard}},
  \bibinfo{author}{\bibfnamefont{C.}~\bibnamefont{Mora}},
  \bibinfo{author}{\bibfnamefont{A.}~\bibnamefont{Cottet}}, \bibnamefont{and}
  \bibinfo{author}{\bibfnamefont{T.}~\bibnamefont{Kontos}}, Coupling a Quantum
  Dot, Fermionic Leads, and a Microwave Cavity on a Chip,
  \bibinfo{journal}{Phys. Rev. Lett.} \textbf{\bibinfo{volume}{107}},
  \bibinfo{pages}{256804} (\bibinfo{year}{2011}).

\bibitem[{\citenamefont{Frey et~al.}(2012)\citenamefont{Frey, Leek, Beck,
  Blais, Ihn, Ensslin, and Wallraff}}]{Frey2012}
\bibinfo{author}{\bibfnamefont{T.}~\bibnamefont{Frey}},
  \bibinfo{author}{\bibfnamefont{P.~J.} \bibnamefont{Leek}},
  \bibinfo{author}{\bibfnamefont{M.}~\bibnamefont{Beck}},
  \bibinfo{author}{\bibfnamefont{A.}~\bibnamefont{Blais}},
  \bibinfo{author}{\bibfnamefont{T.}~\bibnamefont{Ihn}},
  \bibinfo{author}{\bibfnamefont{K.}~\bibnamefont{Ensslin}}, \bibnamefont{and}
  \bibinfo{author}{\bibfnamefont{A.}~\bibnamefont{Wallraff}}, Dipole Coupling
  of a Double Quantum Dot to a Microwave Resonator, \bibinfo{journal}{Phys.
  Rev. Lett.} \textbf{\bibinfo{volume}{108}}, \bibinfo{pages}{046807}
  (\bibinfo{year}{2012}).

\bibitem[{\citenamefont{Mi et~al.}(2017)\citenamefont{Mi, Cady, Zajac, Deelman,
  and Petta}}]{Mi2017}
\bibinfo{author}{\bibfnamefont{X.}~\bibnamefont{Mi}},
  \bibinfo{author}{\bibfnamefont{J.~V.} \bibnamefont{Cady}},
  \bibinfo{author}{\bibfnamefont{D.~M.} \bibnamefont{Zajac}},
  \bibinfo{author}{\bibfnamefont{P.~W.} \bibnamefont{Deelman}},
  \bibnamefont{and} \bibinfo{author}{\bibfnamefont{J.~R.} \bibnamefont{Petta}},
  Strong Coupling of a Single Electron in Silicon to a Microwave Photon,
  \bibinfo{journal}{Science} \textbf{\bibinfo{volume}{355}},
  \bibinfo{pages}{156} (\bibinfo{year}{2017}).

\bibitem[{\citenamefont{{Bruhat} et~al.}(2016)\citenamefont{{Bruhat},
  {Cubaynes}, {Viennot}, {Dartiailh}, {Desjardins}, {Cottet}, and
  {Kontos}}}]{Bruhat2016}
\bibinfo{author}{\bibfnamefont{L.~E.} \bibnamefont{{Bruhat}}},
  \bibinfo{author}{\bibfnamefont{T.}~\bibnamefont{{Cubaynes}}},
  \bibinfo{author}{\bibfnamefont{J.~J.} \bibnamefont{{Viennot}}},
  \bibinfo{author}{\bibfnamefont{M.~C.} \bibnamefont{{Dartiailh}}},
  \bibinfo{author}{\bibfnamefont{M.~M.} \bibnamefont{{Desjardins}}},
  \bibinfo{author}{\bibfnamefont{A.}~\bibnamefont{{Cottet}}}, \bibnamefont{and}
  \bibinfo{author}{\bibfnamefont{T.}~\bibnamefont{{Kontos}}}, Strong Coupling
  Between an Electron in a Quantum Dot Circuit and a Photon in a Cavity,
  \bibinfo{journal}{arxiv: 1612.05214}  (\bibinfo{year}{2016}).

\bibitem[{\citenamefont{Stockklauser et~al.}(2017)\citenamefont{Stockklauser,
  Scarlino, Koski, Gasparinetti, Andersen, Reichl, Wegscheider, Ihn, Ensslin,
  and Wallraff}}]{Stockklauser2017}
\bibinfo{author}{\bibfnamefont{A.}~\bibnamefont{Stockklauser}},
  \bibinfo{author}{\bibfnamefont{P.}~\bibnamefont{Scarlino}},
  \bibinfo{author}{\bibfnamefont{J.~V.} \bibnamefont{Koski}},
  \bibinfo{author}{\bibfnamefont{S.}~\bibnamefont{Gasparinetti}},
  \bibinfo{author}{\bibfnamefont{C.~K.} \bibnamefont{Andersen}},
  \bibinfo{author}{\bibfnamefont{C.}~\bibnamefont{Reichl}},
  \bibinfo{author}{\bibfnamefont{W.}~\bibnamefont{Wegscheider}},
  \bibinfo{author}{\bibfnamefont{T.}~\bibnamefont{Ihn}},
  \bibinfo{author}{\bibfnamefont{K.}~\bibnamefont{Ensslin}}, \bibnamefont{and}
  \bibinfo{author}{\bibfnamefont{A.}~\bibnamefont{Wallraff}}, Strong Coupling
  Cavity QED with Gate-Defined Double Quantum Dots Enabled by a High Impedance
  Resonator, \bibinfo{journal}{Phys. Rev. X} \textbf{\bibinfo{volume}{7}},
  \bibinfo{pages}{011030} (\bibinfo{year}{2017}).

\bibitem[{\citenamefont{Astafiev et~al.}(2007)\citenamefont{Astafiev, Inomata,
  Niskanen, Yamamoto, Pashkin, Nakamura, and Tsai}}]{Astafiev2007}
\bibinfo{author}{\bibfnamefont{O.}~\bibnamefont{Astafiev}},
  \bibinfo{author}{\bibfnamefont{K.}~\bibnamefont{Inomata}},
  \bibinfo{author}{\bibfnamefont{A.~O.} \bibnamefont{Niskanen}},
  \bibinfo{author}{\bibfnamefont{T.}~\bibnamefont{Yamamoto}},
  \bibinfo{author}{\bibfnamefont{Y.~A.} \bibnamefont{Pashkin}},
  \bibinfo{author}{\bibfnamefont{Y.}~\bibnamefont{Nakamura}}, \bibnamefont{and}
  \bibinfo{author}{\bibfnamefont{J.~S.} \bibnamefont{Tsai}}, Single
  Artificial-Atom Lasing, \bibinfo{journal}{Nature (London)}
  \textbf{\bibinfo{volume}{449}}, \bibinfo{pages}{588} (\bibinfo{year}{2007}).

\bibitem[{\citenamefont{Cassidy et~al.}(2017)\citenamefont{Cassidy, Bruno,
  Rubbert, Irfan, Kammhuber, Schouten, Akhmerov, and
  Kouwenhoven}}]{Cassidy2017}
\bibinfo{author}{\bibfnamefont{M.~C.} \bibnamefont{Cassidy}},
  \bibinfo{author}{\bibfnamefont{A.}~\bibnamefont{Bruno}},
  \bibinfo{author}{\bibfnamefont{S.}~\bibnamefont{Rubbert}},
  \bibinfo{author}{\bibfnamefont{M.}~\bibnamefont{Irfan}},
  \bibinfo{author}{\bibfnamefont{J.}~\bibnamefont{Kammhuber}},
  \bibinfo{author}{\bibfnamefont{R.~N.} \bibnamefont{Schouten}},
  \bibinfo{author}{\bibfnamefont{A.~R.} \bibnamefont{Akhmerov}},
  \bibnamefont{and} \bibinfo{author}{\bibfnamefont{L.~P.}
  \bibnamefont{Kouwenhoven}}, Demonstration of an ac Josephson junction laser,
  \bibinfo{journal}{Science} \textbf{\bibinfo{volume}{355}},
  \bibinfo{pages}{939} (\bibinfo{year}{2017}).

\bibitem[{\citenamefont{Stockklauser et~al.}(2015)\citenamefont{Stockklauser,
  Maisi, Basset, Cujia, Reichl, Wegscheider, Ihn, Wallraff, and
  Ensslin}}]{Stockklauser2015}
\bibinfo{author}{\bibfnamefont{A.}~\bibnamefont{Stockklauser}},
  \bibinfo{author}{\bibfnamefont{V.~F.} \bibnamefont{Maisi}},
  \bibinfo{author}{\bibfnamefont{J.}~\bibnamefont{Basset}},
  \bibinfo{author}{\bibfnamefont{K.}~\bibnamefont{Cujia}},
  \bibinfo{author}{\bibfnamefont{C.}~\bibnamefont{Reichl}},
  \bibinfo{author}{\bibfnamefont{W.}~\bibnamefont{Wegscheider}},
  \bibinfo{author}{\bibfnamefont{T.}~\bibnamefont{Ihn}},
  \bibinfo{author}{\bibfnamefont{A.}~\bibnamefont{Wallraff}}, \bibnamefont{and}
  \bibinfo{author}{\bibfnamefont{K.}~\bibnamefont{Ensslin}}, Microwave Emission
  from Hybridized States in a Semiconductor Charge Qubit,
  \bibinfo{journal}{Phys. Rev. Lett.} \textbf{\bibinfo{volume}{115}},
  \bibinfo{pages}{046802} (\bibinfo{year}{2015}).

\bibitem[{\citenamefont{Stehlik et~al.}(2016)\citenamefont{Stehlik, Liu,
  Eichler, Hartke, Mi, Gullans, Taylor, and Petta}}]{Stehlik2016}
\bibinfo{author}{\bibfnamefont{J.}~\bibnamefont{Stehlik}},
  \bibinfo{author}{\bibfnamefont{Y.-Y.} \bibnamefont{Liu}},
  \bibinfo{author}{\bibfnamefont{C.}~\bibnamefont{Eichler}},
  \bibinfo{author}{\bibfnamefont{T.~R.} \bibnamefont{Hartke}},
  \bibinfo{author}{\bibfnamefont{X.}~\bibnamefont{Mi}},
  \bibinfo{author}{\bibfnamefont{M.~J.} \bibnamefont{Gullans}},
  \bibinfo{author}{\bibfnamefont{J.~M.} \bibnamefont{Taylor}},
  \bibnamefont{and} \bibinfo{author}{\bibfnamefont{J.~R.} \bibnamefont{Petta}},
  Double Quantum Dot Floquet Gain Medium, \bibinfo{journal}{Phys. Rev. X}
  \textbf{\bibinfo{volume}{6}}, \bibinfo{pages}{041027} (\bibinfo{year}{2016}).

\bibitem[{\citenamefont{{Liu} et~al.}(2017)\citenamefont{{Liu}, {Stehlik},
  {Eichler}, {Mi}, {Hartke}, {Gullans}, {Taylor}, and {Petta}}}]{Liu2017}
\bibinfo{author}{\bibfnamefont{Y.-Y.} \bibnamefont{{Liu}}},
  \bibinfo{author}{\bibfnamefont{J.}~\bibnamefont{{Stehlik}}},
  \bibinfo{author}{\bibfnamefont{C.}~\bibnamefont{{Eichler}}},
  \bibinfo{author}{\bibfnamefont{X.}~\bibnamefont{{Mi}}},
  \bibinfo{author}{\bibfnamefont{T.}~\bibnamefont{{Hartke}}},
  \bibinfo{author}{\bibfnamefont{M.~J.} \bibnamefont{{Gullans}}},
  \bibinfo{author}{\bibfnamefont{J.~M.} \bibnamefont{{Taylor}}},
  \bibnamefont{and} \bibinfo{author}{\bibfnamefont{J.~R.}
  \bibnamefont{{Petta}}}, {Threshold Dynamics of a Semiconductor Single Atom
  Maser}, \bibinfo{journal}{arxiv: 1704.01961}  (\bibinfo{year}{2017}).

\bibitem[{\citenamefont{van~den Berg et~al.}(2014)\citenamefont{van~den Berg,
  Bergenfeldt, and Samuelsson}}]{Berg2014}
\bibinfo{author}{\bibfnamefont{T.~L.} \bibnamefont{van~den Berg}},
  \bibinfo{author}{\bibfnamefont{C.}~\bibnamefont{Bergenfeldt}},
  \bibnamefont{and}
  \bibinfo{author}{\bibfnamefont{P.}~\bibnamefont{Samuelsson}}, Pump-probe
  scheme for electron-photon dynamics in hybrid conductor-cavity systems,
  \bibinfo{journal}{Phys. Rev. B} \textbf{\bibinfo{volume}{90}},
  \bibinfo{pages}{085416} (\bibinfo{year}{2014}).

\bibitem[{\citenamefont{Liu et~al.}(2015{\natexlab{a}})\citenamefont{Liu,
  Stehlik, Eichler, Gullans, Taylor, and Petta}}]{Liu2015}
\bibinfo{author}{\bibfnamefont{Y.-Y.} \bibnamefont{Liu}},
  \bibinfo{author}{\bibfnamefont{J.}~\bibnamefont{Stehlik}},
  \bibinfo{author}{\bibfnamefont{C.}~\bibnamefont{Eichler}},
  \bibinfo{author}{\bibfnamefont{M.~J.} \bibnamefont{Gullans}},
  \bibinfo{author}{\bibfnamefont{J.~M.} \bibnamefont{Taylor}},
  \bibnamefont{and} \bibinfo{author}{\bibfnamefont{J.~R.} \bibnamefont{Petta}},
  Semiconductor Double Quantum Dot Micromaser, \bibinfo{journal}{Science}
  \textbf{\bibinfo{volume}{347}}, \bibinfo{pages}{285}
  (\bibinfo{year}{2015}{\natexlab{a}}).

\bibitem[{\citenamefont{Eichler et~al.}(2012)\citenamefont{Eichler, Bozyigit,
  and Wallraff}}]{EichlerPRA2012}
\bibinfo{author}{\bibfnamefont{C.}~\bibnamefont{Eichler}},
  \bibinfo{author}{\bibfnamefont{D.}~\bibnamefont{Bozyigit}}, \bibnamefont{and}
  \bibinfo{author}{\bibfnamefont{A.}~\bibnamefont{Wallraff}}, Characterizing
  Quantum Microwave Radiation and its Entanglement with Superconducting Qubits
  Using Linear Detectors, \bibinfo{journal}{Phys. Rev. A}
  \textbf{\bibinfo{volume}{86}}, \bibinfo{pages}{032106}
  (\bibinfo{year}{2012}).

\bibitem[{\citenamefont{Liu et~al.}(2015{\natexlab{b}})\citenamefont{Liu,
  Stehlik, Gullans, Taylor, and Petta}}]{LiuPRA2015}
\bibinfo{author}{\bibfnamefont{Y.-Y.} \bibnamefont{Liu}},
  \bibinfo{author}{\bibfnamefont{J.}~\bibnamefont{Stehlik}},
  \bibinfo{author}{\bibfnamefont{M.~J.} \bibnamefont{Gullans}},
  \bibinfo{author}{\bibfnamefont{J.~M.} \bibnamefont{Taylor}},
  \bibnamefont{and} \bibinfo{author}{\bibfnamefont{J.~R.} \bibnamefont{Petta}},
  Injection locking of a semiconductor double-quantum-dot micromaser,
  \bibinfo{journal}{Phys. Rev. A} \textbf{\bibinfo{volume}{92}},
  \bibinfo{pages}{053802} (\bibinfo{year}{2015}{\natexlab{b}}).

\bibitem[{\citenamefont{Reed et~al.}(2010)\citenamefont{Reed, DiCarlo, Johnson,
  Sun, Schuster, Frunzio, and Schoelkopf}}]{Reed2010}
\bibinfo{author}{\bibfnamefont{M.~D.} \bibnamefont{Reed}},
  \bibinfo{author}{\bibfnamefont{L.}~\bibnamefont{DiCarlo}},
  \bibinfo{author}{\bibfnamefont{B.~R.} \bibnamefont{Johnson}},
  \bibinfo{author}{\bibfnamefont{L.}~\bibnamefont{Sun}},
  \bibinfo{author}{\bibfnamefont{D.~I.} \bibnamefont{Schuster}},
  \bibinfo{author}{\bibfnamefont{L.}~\bibnamefont{Frunzio}}, \bibnamefont{and}
  \bibinfo{author}{\bibfnamefont{R.~J.} \bibnamefont{Schoelkopf}},
  High-Fidelity Readout in Circuit Quantum Electrodynamics Using the
  Jaynes-Cummings Nonlinearity, \bibinfo{journal}{Phys. Rev. Lett.}
  \textbf{\bibinfo{volume}{105}}, \bibinfo{pages}{173601}
  (\bibinfo{year}{2010}).

\end{thebibliography}

\end{document}